\documentclass[aps,pra,twocolumn,superscriptaddress,amsmath,amssymb]{revtex4}
\usepackage{graphicx}
\usepackage{amsmath}
\usepackage{color}
\usepackage[normalem]{ulem}
\def\cp#1{\mathbf{#1}}

\begin{document}

\title{Quench dynamics of a Bose-Einstein condensate under synthetic spin-orbit coupling}

\author{Tian-Shu Deng}
\affiliation{Key Laboratory of Quantum Information, University of Science and Technology of China,
CAS, Hefei, Anhui, 230026, People's Republic of China}
\affiliation{Synergetic Innovation Center of Quantum Information and Quantum Physics, University of Science and Technology of China, Hefei, Anhui 230026, China}
\author{Wei Zhang}
\email{wzhangl@ruc.edu.cn}
\affiliation{Department of Physics, Renmin University of China, Beijing 100872, China}
\affiliation{Beijing Key Laboratory of Opto-electronic Functional Materials and Micro-nano Devices,
Renmin University of China, Beijing 100872, China}
\author{Wei Yi}
\email{wyiz@ustc.edu.cn}
\affiliation{Key Laboratory of Quantum Information, University of Science and Technology of China,
CAS, Hefei, Anhui, 230026, People's Republic of China}
\affiliation{Synergetic Innovation Center of Quantum Information and Quantum Physics, University of Science and Technology of China, Hefei, Anhui 230026, China}
\author{Guang-Can Guo}
\affiliation{Key Laboratory of Quantum Information, University of Science and Technology of China,
CAS, Hefei, Anhui, 230026, People's Republic of China}
\affiliation{Synergetic Innovation Center of Quantum Information and Quantum Physics, University of Science and Technology of China, Hefei, Anhui 230026, China}

\date{\today}
\begin{abstract}
We study the quench dynamics of a Bose-Einstein condensate under a Raman-assisted synthetic spin-orbit coupling. To model the dynamical process, we adopt a self-consistent Bogoliubov approach, which is equivalent to applying the time-dependent Bogoliubov-de-Gennes equations. We investigate the dynamics of the condensate fraction as well as the momentum distribution of the Bose gas following a sudden change of system parameters. Typically, the system evolves into a steady state in the long-time limit, which features an oscillating momentum distribution and a stationary condensate fraction. We investigate how different quench parameters such as the inter- and intra-species interactions and the spin-orbit-coupling parameters affect the condensate fraction in the steady state. Furthermore, we find that the time average of the oscillatory momentum distribution in the long-time limit can be described by a generalized Gibbs ensemble with two branches of momentum-dependent Gibbs temperatures. Our study is relevant to the experimental investigation of dynamical processes in a spin-orbit coupled Bose-Einstein condensate.
\end{abstract}
\maketitle

\section{Introduction}
The study of quantum quench, the evolution following a sudden or slow change of the coupling constants of the system Hamiltonian, has attracted much attention in recent years. Dynamics in quench processes can serve as a powerful tool in revealing rich correlations and inducing interesting non-equilibrium states~\cite{quenchrmp1,quenchrmp2}. Due to the highly controllable parameters, ultracold atomic gases have become an ideal platform for the investigation of dynamical processes~\cite{quenchrmp1}. Experimentally, quench dynamics have been used to study various interesting non-equilibrium properties of a Bose-Einstein condensate (BEC)~\cite{boseexp1,boseexp2,boseexp3,boseexp4,boseexp5}, and quenching has proved pivotal in probing pairing correlations in Fermi condensates~\cite{fermiexp1,fermiexp2}. Theoretically, quench dynamics of both BEC and Fermi superfluid have been extensively investigated, leading to predictions of various novel steady-state phases and phase transitions~\cite{quenchtheory1,quenchtheory2,quenchtheory3,quenchtheory4,quenchtheory5,quenchtheory6, quenchtheory7,quenchtheory8,quenchtheory9}. As the ground-state phase and many-body correlations depend sensitively on the inter-atomic interaction, a frequent theme in these studies is investigating the system dynamics following a sudden change of inter-atomic interaction strength. This is possible by switching the external magnetic field via a Feshbach resonance~\cite{feshreview}.

A potential problem in these quench processes is the large atom losses close to the Feshbach resonance. This is less of a problem for Fermi gases, where losses through three-body recombination are suppressed near resonance~\cite{lossfermi}. For Bose gases, the three-body loss rate typically grows faster with an increasing scattering length than the two-body scattering rate~\cite{lossbose}. As the two-body scattering is essential for equilibration, it has been difficult to realize a strongly interacting BEC in equilibrium close to the Feshbach resonance. However, as a recent JILA experiment demonstrates~\cite{jinexp}, close to unitarity, a steady-state BEC with strong interactions can be prepared via a quench process, where saturation of the system properties have been observed before the atoms are lost due to three-body recombination. Apparently, the three-body loss rate is lower than the two-body scattering rate when the system is quenched close to unitarity. This is not unexpected, as at unitarity, the scattering length diverges and the scaling relations of losses with respect to the scattering length are no longer applicable~\cite{jinexp}. This allows the gas to evolve into a stead-state BEC. Theoretically, the quench process and the final steady-state properties can be qualitatively captured by dynamic mean field approaches~\cite{rad,HYL}.

Motivated by these exciting experimental and theoretical developments, we study the quench dynamics of a BEC under the synthetic spin-orbit coupling (SOC) that has recently been realized experimentally at the National Institute of Standards and Technology (NIST)~\cite{gauge2exp,fermisocexp1,fermisocexp2}. The most important effect of SOC on the system is the modification of the single-particle dispersion~\cite{socreview1,socreview2,socreview3,socreview4,socreview5,socreview6,socreview7}, which can give rise to interesting phases and phase transitions. For a spin-orbit coupled Bose gas, quench processes have already been implemented experimentally to study collective behavior such as dipole oscillations and atom losses in a potential trap~\cite{shuaiexp1,shuaiexp2}.

In this work, we focus on the quench dynamics of a spin-orbit coupled BEC. To model the dynamical process, we adopt a self-consistent Bogoliubov approach, which is equivalent to applying the time-dependent Bogoliubov-de-Gennes equations. We investigate the dynamics of the condensate fraction as well as the momentum distribution of the Bose gas after a sudden change of system parameters such as the interaction strength or the laser parameters generating the SOC. Typically, the system evolves into a steady state in the long-time limit, which features a stationary condensate fraction and an oscillating momentum distribution. The condensate fraction of the steady state depends on the quench parameters and remains finite. This is consistent with a recent work on the quench dynamics of a BEC in the absence of SOC~\cite{HYL}. Although our approach differs from the approach in Ref.~\cite{HYL}, one thing in common is that the variation of the condensate mean field in time is taken into account. We also find that while sudden changes of SU(2)-invariant interaction strengths can lead to large condensate depletions, a sudden change of the inter-species interaction strength typically gives rise to a much smaller condensate depletion. Therefore, the intra-species interactions have great impact on the steady-state condensate fraction in these system. Furthermore, we show that if the SOC strength is changed, which is feasible by controlling the laser strength, the condensate fraction would remain large in the dynamical process. Surprisingly, the depletion of the condensate remains small even if the single-particle dispersion undergoes a qualitative change after the quench. Finally, we find that the time average of the oscillatory momentum distribution in the long-time limit can be described by a generalized Gibbs ensemble with two branches of momentum-dependent Gibbs temperatures, which correspond to the two helicity branches in the Bogoliubov excitation spectrum. Our study is relevant to the experimental investigation of dynamical processes in a spin-orbit coupled Bose-Einstein condensate.

The paper is organized as follows. In Sec. II, we present the model that we use to simulate the quench dynamics. We then discuss our main results in Sec. III. In Sec. IV, we resort to the generalized Gibbs ensemble to describe the momentum distribution of the steady state in the long time limit. Finally, we summarize in Sec. V.

\section{Model}

We consider a two-component BEC in three spatial dimensions, which is subject to the one-dimensional SOC that has recently realized at NIST. The single-particle Hamiltonian of the system can be written as~\cite{jpb}
\begin{equation}
    \hat{H}_{0}=\left[\begin{array}{cc}
\frac{(\cp p+\hbar k_{r}\cp e_{x})^{2}}{2m}+\frac{\delta}{2} & \frac{\Omega}{2}\\
\frac{\Omega}{2} & \frac{(\cp p-\hbar k_{r}\cp e_{x})^{2}}{2m}-\frac{\delta}{2}
\end{array}\right],
\end{equation}
where $\Omega$ is the effective Rabi frequency of the Raman process generating the SOC, $k_r$ is the recoil momentum, and $\delta$ is the two-photon detuning of the Raman process, which we take to be zero throughout the work for simplicity. It would be straightforward to generalize out approach to other cases. Here, we have taken the basis $\{\psi_{\uparrow}(\cp r),\psi_{\downarrow}(\cp r)\}^T$, where $\psi_{\sigma}$ ($\sigma=\uparrow,\downarrow$) is the field operator for the corresponding spin species. The SOC mixes different spin species into the so-called helicity branches, and modifies the single-particle dispersion along the $x$-direction. The single-particle dispersion for the two helcity branches can be given as
\begin{equation}
\varepsilon_{\cp k,\pm}=\frac{k^{2}}{2m}+\frac{k_{r}^{2}}{2m}\pm\sqrt{\left(\frac{{k_{x}}{k_{r}}}{m}-\frac{\delta}{2}\right)^{2}+\frac{\Omega^2}{4}},
\label{eqn:helicity}
\end{equation}
where we take $\hbar = 1$ to simplify notation. The lower branch of the single-particle dispersion $\varepsilon_-$ has a double-well structure along the $x$-direction when $\Omega<4E_r$; while for $\Omega>4E_r$, $\varepsilon_-$ only has a single minimum along the $x$-direction. Here, the recoil energy $E_r= k_r^2/2m$.

The Hamiltonian describing inter-atomic interactions can be written as
\begin{eqnarray}
    \hat{H}_{int}&=&\int d^{3}\cp r \Big(\frac{1}{2}g_{1}\psi_{\uparrow}^{\dagger}\psi_{\uparrow}^{\dagger}\psi_{\uparrow} \psi_{\uparrow}
    \nonumber \\
    &&+\frac{1}{2}g_{2}\psi_{\downarrow}^{\dagger}\psi_{\downarrow}^{\dagger}\psi_{\downarrow} \psi_{\downarrow}+g_{12}\psi_{\uparrow}^{\dagger}\psi_{\downarrow}^{\dagger}\psi_{\downarrow}\psi_{\uparrow} \Big),
\end{eqnarray}
where $g_i$ ($i=1,2$) is the intra-species interaction rate and $g_{12}$ gives the inter-species interaction. Both the ground-state phases and the excitation spectrum of this system have been extensively studied previously. Importantly, the ground state can be either a plane-wave phase or a stripe phase, depending on the interaction parameters~\cite{jpb}. In this work, for simplicity, we assume that initially the system is in a plane-wave phase with an SU(2) invariant interactions, i.e., $g_1=g_2=g_{12}=g$. For the dynamical process, we consider either changing the interactions simultaneously while maintaining the SU(2) symmetry, or changing the inter-species interaction $g_{12}$.

Following the common practice, we write the field operators as
\begin{equation}
\left[\begin{array}{c}
\psi_{\uparrow}(\cp r,t)\\
\psi_{\downarrow}(\cp r,t)
\end{array}\right]=\left[\begin{array}{c}
\varphi_{\uparrow}(\cp r)\\
\varphi_{\text{\ensuremath{\downarrow}}}(\cp r)
\end{array}\right]e^{-i\mu t/\hbar}+\left[\begin{array}{c}
\delta \psi_{\uparrow}(\cp r,t)\\
\delta \psi_{\downarrow}(\cp r,t)
\end{array}\right],
\end{equation}
where $\varphi_{\sigma}(\cp r)$ are the wave functions of the equilibrium ground state, $\delta\psi_{\sigma}$ are field operators associated with fluctuations, and $\mu$ is the chemical potential.

The ground-state wave function satisfies the Gross-Pitaevskii (GP) equation
\begin{align}
&\left[\hat{H}_{0}+\left(\begin{array}{cc}
g|\varphi_{\uparrow}|^{2}+g_{12}|\varphi_{\downarrow}|^{2} & 0\\
0 & g_{12}|\varphi_{\uparrow}|^{2}+g|\varphi_{\downarrow}|^{2}
\end{array}\right)\right]\left[\begin{array}{c}
\varphi_{\uparrow}(\cp{r})\\
\varphi_{\downarrow}(\cp{r})
\end{array}\right]\nonumber\\
&=\mu\left[\begin{array}{c}
\varphi_{\uparrow}(\cp{r})\\
\varphi_{\downarrow}(\cp{r})
\end{array}\right],
\end{align}
where we have taken $g_1=g_2=g$. Under the one-dimensional SOC that we are considering here, the wave function can be written as
\begin{equation}
\left[\begin{array}{c}
\varphi_{\uparrow}(\cp r)\\
\varphi_{\text{\ensuremath{\downarrow}}}(\cp r)
\end{array}\right]=\sqrt{n_{0}}\left[\begin{array}{c}
\cos\theta\\
-\sin\theta
\end{array}\right]\exp(ip_{0}x),
\end{equation}
where $n_0$ is the condensate fraction, $p_0$ denotes the condensation momentum in the plane-wave state, and $\theta$ is typically spatially independent. For initial states with SU(2)-invariant interaction, $p_0$ can be determined by numerically minimizing the single-particle dispersion of the lower branch in Eq.~\ref{eqn:helicity}~\cite{jpb}. The parameter $\theta$ as well as the chemical potential $\mu$ can be determined from the GP equation above.

As we are only considering the homogeneous case, it is convenient to examine the fluctuations in momentum space. After taking the Fourier transform $\delta\psi_{\sigma}=\frac{1}{\sqrt{V}}\sum_{\cp k}\exp(i\cp k\cdot\cp r)\hat{b}_{\cp k,\sigma}$, the total Hamiltonian becomes
 \begin{equation}
    \hat{H}=\frac{1}{2}\sum_{\cp k}\hat{B}_{\cp k}^{\dag}H_{\cp k}\hat{B}_{\cp k}+E_{0},
 \end{equation}
where $E_0$ is a constant energy shift with no effects on dynamical processes, $\hat{B}^{\dag}_{\cp k}=\left\{\hat{b}_{\cp q+\cp k,\uparrow}^{\dag},\hat{b}_{\cp q+\cp k,\downarrow}^{\dag}, \hat{b}_{\cp q-\cp k,\uparrow},\hat{b}_{\cp q-\cp k,\downarrow}\right\}$, $\cp q=p_0 \cp e_{x}$, and
 \begin{equation}
    H_{k}=\left[\begin{array}{cc}
H_{0}(\cp k+\cp q)+A-{\mu}I & B\\
B & H_{0}(-\cp k+\cp q)+A-{\mu}I
\end{array}\right].
 \end{equation}
Here,
\begin{widetext}
 \begin{align}
A=\left[\begin{array}{cc}
(2gn_{0}\cos^{2}\theta+g_{12}n_{0}\sin^{2}\theta) & -g_{12}n_{0}\sin\theta\cos\theta\\
-g_{12}n_{0}\sin\theta\cos\theta & (2gn_{0}\sin^{2}\theta+g_{12}n_{0}\cos^{2}\theta)
\end{array}\right],\quad
B=\left(\begin{array}{cc}
gn_{0}cos^{2}\theta & -g_{12}n_{0}\sin\theta\cos\theta\\
-g_{12}n_{0}\sin\theta\cos\theta & gn_{0}\sin^{2}\theta
\end{array}\right).
\end{align}
\end{widetext}

We study the quench process where the system in equilibrium at $t=0^-$ undergoes a sudden change of parameters at $t=0$. The system is then left to evolve under the post-quench parameters for $t>0$. The Heisenberg equation for the post-quench operators in momentum space satisfies
\begin{equation}
 \label{h}
    i\tau_{z}\frac{\partial}{\partial t}\left(\begin{array}{c}
b_{\cp q+\cp k,\uparrow}\\
b_{\cp q+\cp k,\downarrow}\\
b_{\cp q-\cp k,\uparrow}^{\dagger}\\
b_{\cp q-\cp k,\downarrow}^{\dagger}
\end{array}\right)=H_{\cp k}\left(\begin{array}{c}
b_{\cp q+\cp k,\uparrow}\\
b{}_{\cp q+\cp k,\downarrow}\\
b_{\cp q-\cp k,\uparrow}^{\dagger}\\
b_{\cp q-\cp k,\downarrow}^{\dagger}
\end{array}\right),
 \end{equation}
where
\begin{equation}
\tau_{z}=\left(\begin{array}{cccc}
1 & 0 & 0 & 0\\
0 & 1 & 0 & 0\\
0 & 0 & -1 & 0\\
0 & 0 & 0 & -1
\end{array}\right).
\end{equation}
In Eq.~\ref{h}, the condensate density $n_0$ appearing in $H_{\cp k}$ should vary with time, which can be self-consistently determined as we will detail below. We further assume that the quasi-particles at any given time during the dynamical process can still be described by a Bogoliubov theory above the condensate at $t=0^-$. This treatment is equivalent to a self-consistent time-dependent Bogoliubov theory, which should provide a qualitatively correct picture when the condensate depletion is not too large. Similar approaches have been adopted to describe quench processes in BEC and Fermi systems~\cite{quenchtheory2,quenchtheory5,quenchtheory7,quenchtheory8,quenchtheory9,jinexp,rad,HYL}. We note that our assumption here implies that $\theta$ is not time dependent.

In this spirit, we define the instantaneous Bogoliubov quasi-particle operators $\{\beta_{\cp q+\cp k,\uparrow}(t), \beta_{\cp q+\cp k,\downarrow}(t), \beta^{\dag}_{\cp q-\cp k,\uparrow}(t), \beta^{\dag}_{\cp q-\cp k,\downarrow}(t)\}$, which diagonalize the time-dependent Hamiltonian $H_{\cp k}(t)$ in Eq.~(\ref{h}) at any given $t>0$. Therefore, we may write
\begin{equation}
 \label{b}
    \left[\begin{array}{c}
b_{\cp q+\cp k,\uparrow}(t)\\
b_{\cp q+\cp k,\downarrow}(t)\\
b_{\cp q-\cp k,\uparrow}^{\dagger}(t)\\
b_{\cp q-\cp k,\downarrow}^{\dagger}(t)
\end{array}\right]=U(t)\left[\begin{array}{c}
\beta_{\cp q+\cp k,\uparrow}(t)\\
\beta_{\cp q+\cp k,\downarrow}(t)\\
\beta_{\cp q-\cp k,\uparrow}^{\dagger}(t)\\
\beta_{\cp q-\cp k,\downarrow}^{\dagger}(t)
\end{array}\right],
 \end{equation}
where the time-dependent Bogoliubov transformation matrix $U(t)$ satisfies
\begin{equation}
 \label{d}
    i\tau_{z}\frac{\partial}{\partial t}U(t)=H_{\cp k}(t)U(t)-\tau_{z}U(t)\tilde{E}(t)\tau_{z}
 \end{equation}
with
\begin{equation}
    \tilde{E}(t)=\left[\begin{array}{cccc}
E_1(t) & 0 & 0 & 0\\
0 & E_2(t) & 0 & 0\\
0 & 0 & E_3(t) & 0\\
0 & 0 & 0 & E_4(t)
\end{array}\right].
\end{equation}
Here $E_i(t)$ ($i=1,2,3,4$) represents the time-dependent Bogoliubov spectrum, which can be computed by diagonalizing the Hamiltonian $H_{\cp k}(t)$ for any given $t>0$.

To further simplify Eq.~\eqref{d}, we define
 \begin{equation}
    \tilde{U}(t)=U(t)\left(\begin{array}{cccc}
e^{-i\int_{0}^{t}E_1dt} & 0 & 0 & 0\\
0 & e^{-i\int_{0}^{t}E_2dt} & 0 & 0\\
0 & 0 & e^{i\int_{0}^{t}E_3dt} & 0\\
0 & 0 & 0 & e^{i\int_{0}^{t}E_4dt}
\end{array}\right),
 \end{equation}
so that Eq.~\eqref{d} becomes
\begin{equation}\label{eqn:timevolve}
    i\tau_{z}\frac{\partial}{\partial t}\tilde{U}(t)=H_{\cp k}(t)\tilde{U}(t).
 \end{equation}

The significance of $\tilde{U}$ is that it relates the atomic operators $\{b_{\cp k,\sigma},b^{\dag}_{\cp k,\sigma}\}$ at any given time after the quench with those of the Bogoliubov quasi-particles just after the quench at $t=0^{+}$, as
 \begin{equation}
    \left[\begin{array}{c}
b_{\cp q+\cp k,\uparrow}(t)\\
b_{\cp q+\cp k,\downarrow}(t)\\
b_{\cp q-\cp k,\uparrow}^{\dagger}(t)\\
b_{\cp q-\cp k,\downarrow}^{\dagger}(t)
\end{array}\right]=\tilde{U}\left[\begin{array}{c}
\beta_{\cp q+\cp k,\uparrow}(0^+)\\
\beta_{\cp q+\cp k,\downarrow}(0^+)\\
\beta_{\cp q-\cp k,\uparrow}^{\dagger}(0^+)\\
\beta_{\cp q-\cp k,\downarrow}^{\dagger}(0^+)
\end{array}\right].
 \end{equation}

As the coupling parameters in the Hamiltonian undergo a sudden change at $t=0$, the pre- and post-quench Bogoliubov quasi-particles become different. However, the atomic operators $\{b_{\cp k,\sigma},b^{\dag}_{\cp k,\sigma}\}$ at $t=0$ are not affected, given a fast enough quench. We may then associate the Bogoliubov quasi-particles before and after the quench using their relations with the atomic operators at $t=0$~\cite{rad}
\begin{equation}
 \left[\begin{array}{c}
\beta_{\cp q+\cp k,\uparrow}(0^+)\\
\beta_{\cp q+\cp k,\downarrow}(0^+)\\
\beta_{\cp q-\cp k,\uparrow}^{\dagger}(0^+)\\
\beta_{\cp q-\cp k,\downarrow}^{\dagger}(0^+)
\end{array}\right]=U(0^+)^{-1}U(0^-)\left[\begin{array}{c}
\beta_{\cp q+\cp k,\uparrow}(0^-)\\
\beta_{\cp q+\cp k,\downarrow}(0^-)\\
\beta_{\cp q-\cp k,\uparrow}^{\dagger}(0^-)\\
\beta_{\cp q-\cp k,\downarrow}^{\dagger}(0^-)
\end{array}\right].
\end{equation}
Experimentally, the condensate fraction is typically measured under the initial pre-quench parameters~\cite{jinexp}. Therefore, the momentum distribution of the atoms at any time during the dynamical process can be calculated by
\begin{align}
   n_{\cp k}(t)&=\langle0-|b_{\cp q+\cp k,\uparrow}^{\dagger} b_{\cp q+\cp k,\uparrow}+ b_{\cp q+\cp k,\text{\ensuremath{\downarrow}}}^{\dagger} b_{\cp q+\cp k,\downarrow}|0-\rangle\nonumber\\
   &=|S_{13}|^{2}+|S_{14}|^{2}+|S_{23}|^{2}+|S_{24}|^{2},\label{eqn:nk}
\end{align}
where
\begin{equation}
    S=\left(\begin{array}{cccc}
S_{11} & S_{12} & S_{13} & S_{14}\\
S_{21} & S_{22} & S_{23} & S_{24}\\
S_{31} & S_{32} & S_{33} & S_{34}\\
S_{41} & S_{42} & S_{43} & S_{44}
\end{array}\right)=\tilde{U}(t)U(0^+)^{-1}U(0^-).
 \end{equation}
The condensate fraction is then
\begin{equation}
n_0(t)=n-\sum_{\cp k} n_{\cp k}.\label{eqn:n0}
\end{equation}
For each time step, we numerically evolve the transformation matrix using Eq.~\eqref{eqn:timevolve}, calculate quantities like the condensate depletion and the momentum distribution of the quasi-particle excitations using Eq.~\eqref{eqn:nk}, and update the condensate fraction using Eq.~\eqref{eqn:n0}. Following this recipe, we determine the time evolution of various quantities numerically.

\section{Numerical results}

In this section, we present the main numerical results. We will first discuss system dynamics under a sudden change of interactions, followed by discussion of changing the SOC parameter.

\subsection{Interaction as the quench parameter}

We first consider the simple case where both the intra- and the inter-particle interaction strengths are changed at the same time, i.e., $g_1=g_2=g_{12}=g$ at all times. In principle, this can be achieved by a combination of optical and magnetic Feshbach resonance~\cite{twofeshbach}. More practically, typical magnetic Feshbach resonances of intra- and inter-particle interactions are located at different magnetic field, it is easy to change either the intra- or the inter-particle interaction. However, as we will show below, it is meaningful to consider the case with SU(2) invariant interactions as most of the key features of the dynamic process can be captured by this simpler scenario.

We start with a weakly interacting BEC under the one-dimensional SOC at equilibrium. With $g_1=g_2=g_{12}=g>0$, the ground state of the system is in a stable plane-wave phase. The many-body ground state can be approximately described by a condensate wave function at one of the two degenerate points in momentum space.

At the time of quench $t=0$, we change the initial interaction strength $g_i=4\pi\hbar^2a_i/m$ at $t=0^-$ to a very large value $g_f=4\pi\hbar^2a_f/m$ close to resonance at $t=0^+$, where $a_i$ ($a_f$) is the initial (final) scattering length. The dynamical quantities are calculated using the theoretical approach outlined in the previous section. For convenience, in the following calculations, we take a typical number density $n/k^3_r=1$. We have checked that all the numerical results are qualitatively similar for experimentally relevant number densities.

\begin{figure}[tbp]
\centering
\includegraphics[width=8cm]{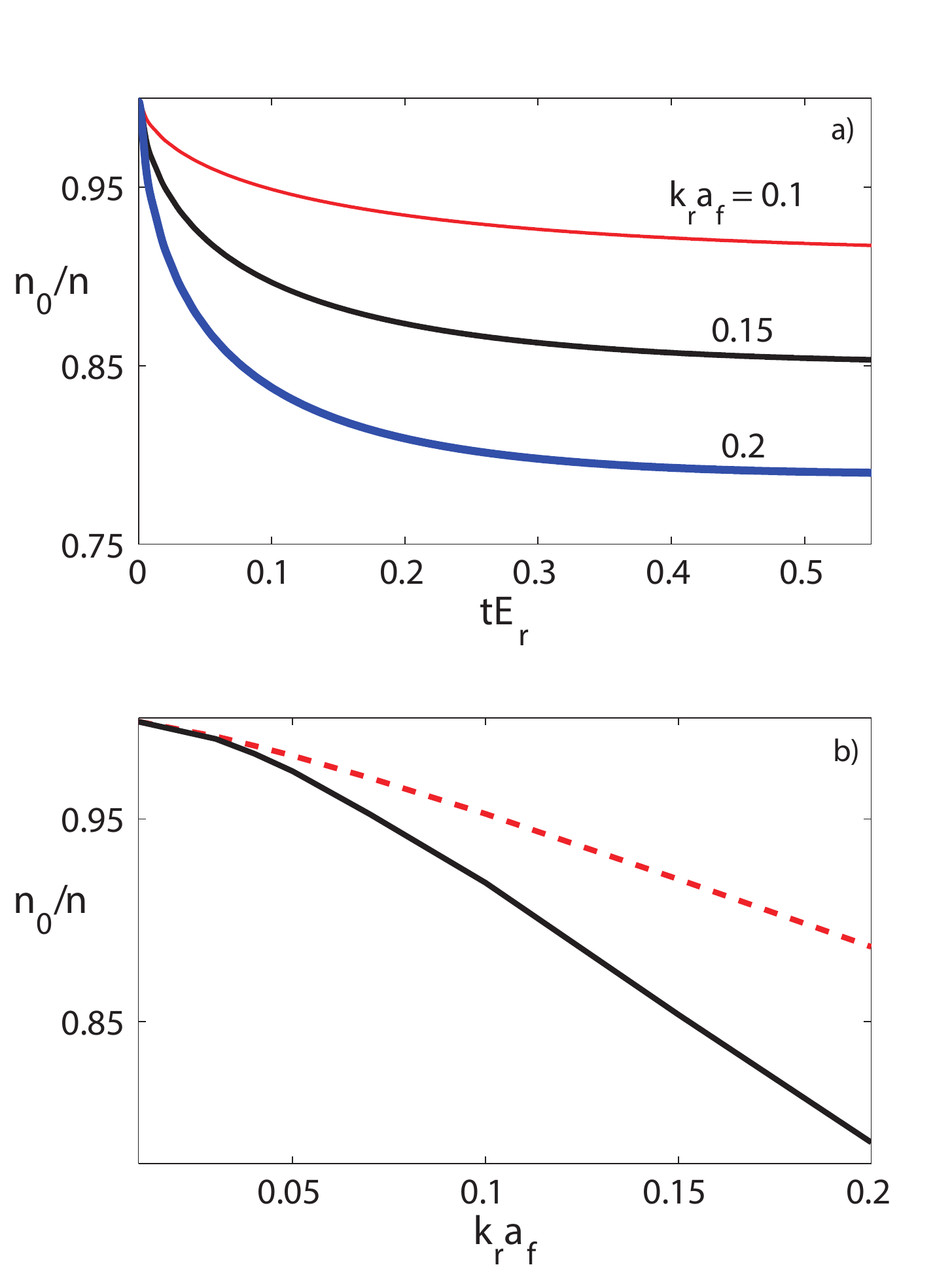}
\caption{(a) Time evolution of the condensate fraction, $n_0/n$ as a function of time $t$, following a sudden change of interaction from a common initial scattering length of $k_ra_i=0.01$ to different final scattering lengths $a_f$. (b) Comparison of the steady-state condensate fraction with that of equilibrium states. The black solid line is the long-time condensate fraction $n_0/n$ as a function of final interaction strength $k_ra_f$, with initial interaction strength $k_ra_i=0.01$. The red dashed line is the condensate fraction in the equilibrium state under the post-quench parameters. Here, the time of evolution $tE_r=0.6$. In both figures, $\Omega/E_r=6$, $\delta=0$.}
\label{fig:fig1}
\end{figure}

Fig.~\ref{fig:fig1}(a) shows the typical post-quench dynamics of the condensate fraction as a function of time. We see that in the long-time limit, regardless of the final interaction strength, the condensate fraction inevitably saturates to a steady-state value, which is much lower than the equilibrium-state value at the final interaction strength. The condensate fraction of the steady state is dependent on the final interaction strength. In Fig.~\ref{fig:fig2}(b), we show the stead-state condensate fraction as a function of the final interaction strength. Apparently, as the final interaction strength increases, the condensate fraction decreases. Importantly, the steady-state condensate fraction remains finite for the largest final interaction strength allowed by our numerical calculation. We note that this observation is consistent with a previous calculation on the quench dynamics of BEC in the absence of SOC~\cite{HYL}. This suggests that the system in the long-time limit would be locked into a steady-state where the condensate fraction saturates to a finite value. This is very different from that of an equilibrium state.

\begin{figure}[tbp]
\centering
\includegraphics[width=8cm]{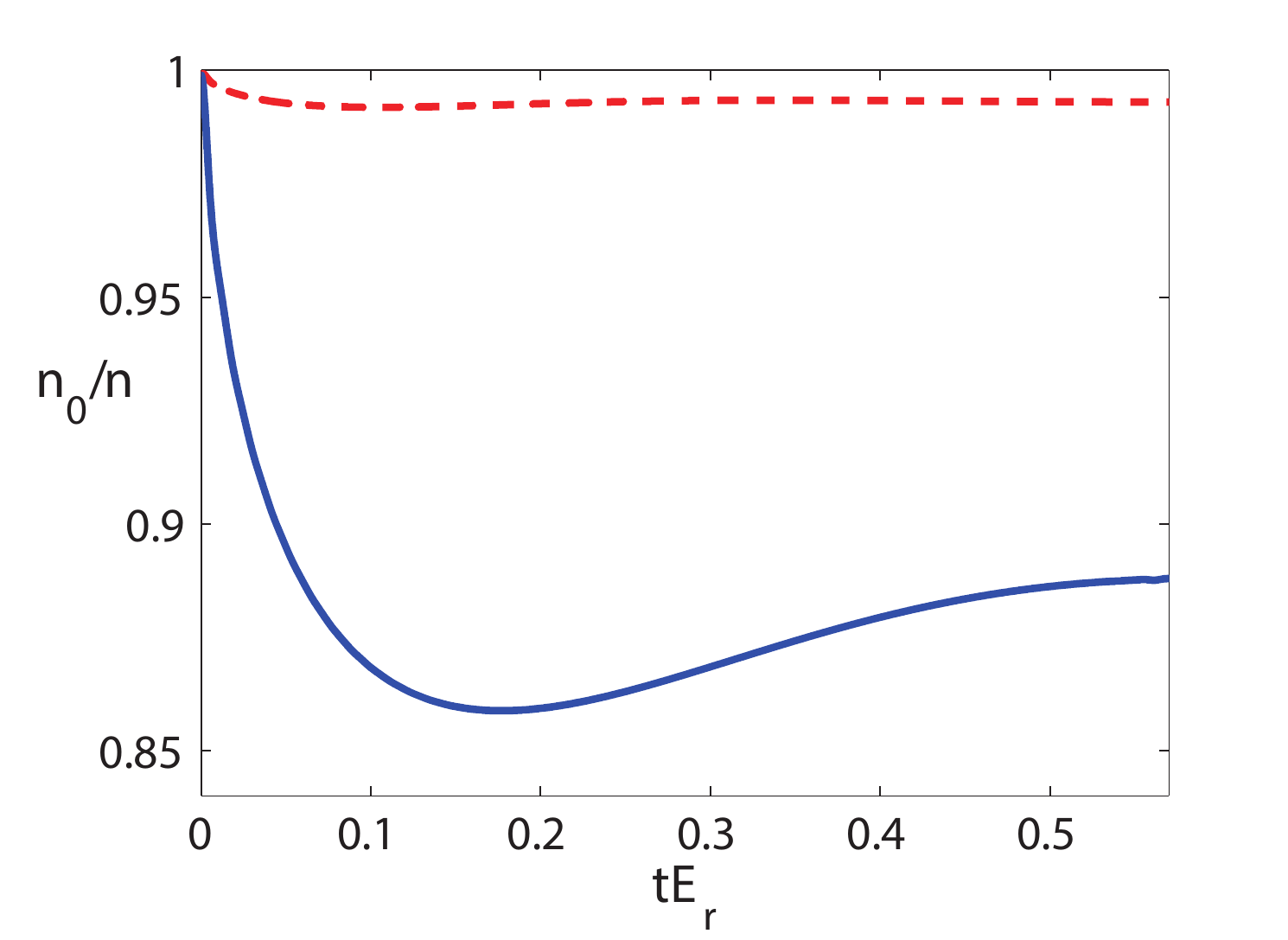}
\caption{Time evolution of the condensate fraction after a sudden change of inter-species interaction. The scattering length $a_i$ associated with the intra-species interaction ($g_1=g_2=g=4\pi\hbar^2a_i/m$) is fixed at $k_ra_i=0.01$, while the scattering length associated with the inter-species interaction $g_{12}$ is changed from $k_ra_i=0.01$ to $k_ra_f=0.6$. For the blue solid line, $\Omega/{E_r}=6$; and for the red dashed line $\Omega/E_r=2$. Here, $\delta=0$.}
\label{fig:ntg12}
\end{figure}

To further understand the role of intra- and inter-species interactions, we also study quench processes where the SU(2) invariance is broken. In Fig.~\ref{fig:ntg12}, we show the time evolution of the condensate fraction when the intra-species interaction is fixed with $g_1=g_2=g=4\pi\hbar^2 a_i/m$, while only the inter-species interaction $g_{12}$ is changed from $g_{12}=4\pi\hbar^2 a_i/m$ to a large value $g_{12}=4\pi\hbar^2 a_f/m$. Typically, the depletion of condensate is much smaller than in the SU(2) invariant case. This demonstrates the critical role of intra-species interaction in generating excitations during the quench process. Finally, we note that quench processes with $\Omega>4E_r$ lead to larger condensate depletions under a sudden change of inter-species interaction. We attribute this to the fact that spin mixing is more pronounced in the pre-quench condensate with $\Omega>4E_r$. Inter-species interaction thus plays a more important role for $\Omega>4E_r$.

\subsection{Changing the SOC parameters}

For a spin-orbit coupled atomic gas, the parameters of the lasers generating the synthetic SOC can also serve as convenient quench parameters. As an example, we study the quench dynamics following a sudden change of the effective Rabi frequency of the Raman process generating the SOC. As the Rabi frequency changes from its initial value $\Omega_i$ to the post-quench value $\Omega_f$, the single-particle dispersion is modified. For $\Omega<4E_r$, the lower helicity branch features a double-well structure, with two degenerate local minima in momentum space. For $\Omega>4E_r$, the lower helicity branch only has one minimum at $k=0$.

\begin{figure}[tbp]
\centering
\includegraphics[width=8cm]{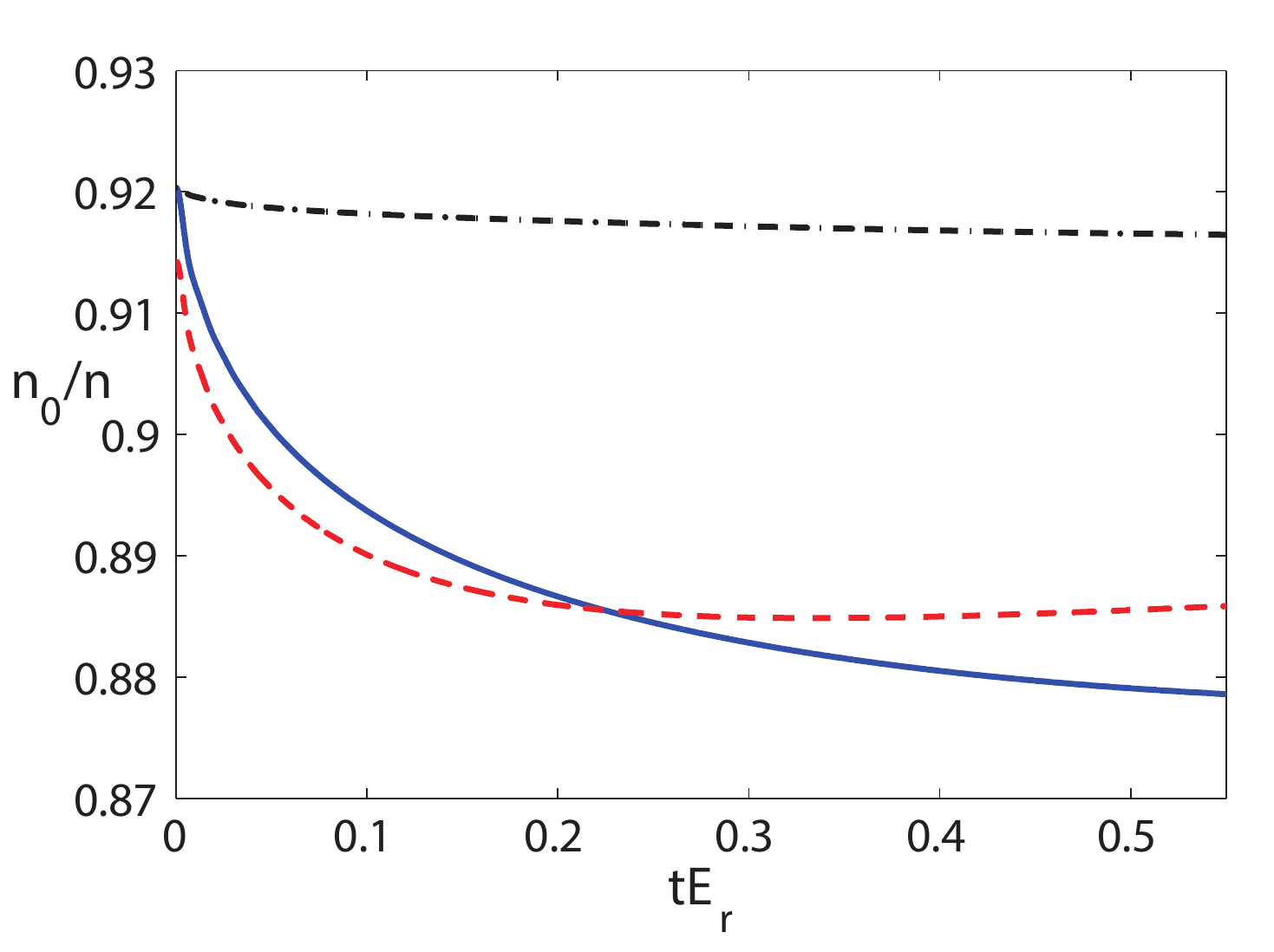}
\caption{Time evolution of the condensate depletion $n_0/n$ following an SOC parameter switch from $\Omega_i$ to $\Omega_f$. The parameters for different quenches are: $\Omega_i/E_r=2$, $\Omega_f/E_r=6$ (red dashed); $\Omega_i/E_r=6$, $\Omega_f/E_r=2$ (blue solid); $\Omega_i/E_r=2$, $\Omega_f/E_r=3$ (black dash-dotted). Here, $g_1=g_2=g_{12}=4\pi \hbar^2 a_s/m$, where the interaction strength is fixed at $k_ra_s=0.16$.}
\label{fig:fig3}
\end{figure}

As demonstrated in Fig.~\ref{fig:fig3}, although the condensate fraction approaches a steady-state value which is different from that of the equilibrium state, the change in the condensate fraction before and after the quench is typically small. This is particularly interesting for the cases where $\Omega_i$ and $\Omega_f$ straddle the critical point $\Omega=4E_r$. In this case, the steady-state condensate condenses at a different state from the equilibrium ground state under post-quench parameters. A similar quasi-condensation has been shown to exist in post-quench steady states of one dimensional systems~\cite{rigol,boseexp5}. Finally, as shown in Fig.~\ref{fig:fig3}, when the critical Rabi frequency $\Omega=4E_r$ is crossed, the modification of the single-particle dispersion leads to a larger change in the steady-state condensate fraction. This is consistent with our expectation, as when both the initial and the final Rabi frequency are on the same side of $4E_r$, the single-particle dispersion around the condensation point in momentum space only undergoes minimal change.

\section{Generalized Gibbs ensemble}

To further understand the steady state, we evoke the formalism of the generalized Gibbs ensemble to characterize the distribution of quasi-particle excitations above the steady-state condensate. It has been pointed out that in isolated integral systems, the steady state can be described by the generalized Gibbs ensemble~\cite{gibbs2}. While in our system the quasi-particle excitations are not isolated due to the existence of a condensate reservoir, we follow the argument first presented in Ref.~\cite{HYL}, that in the long-time limit, the quasi-particle excitations are approximately constants of motion. Thus, the quasi-particles can be treated as an integrable system, and we may describe the system with the generalize Gibbs ensemble.

Following the standard practice, in a generalized Gibbs ensemble, the density operator is given as~\cite{HYL}
\begin{align}
\hat{\rho}_{G}=Z_{G}^{-1}\exp(-\sum_{s}\frac{E_{s}\beta_{s}^{\dag}\beta_{s}}{k_B T_{s}}),
\end{align}
where $Z_{G}={\rm Tr} \exp(-\sum_{s}E_{s}\beta_{s}^{\dag}\beta_{s}/k_B T_{s})$, $E_s$ is the dispersion of the $s$-th mode, $T_s$ is the Gibbs temperature of the $s$-th mode. In the absence of SOC, it has been shown in Ref.~\cite{HYL} that the steady-state in the long-time limit can be well described by the generalized Gibbs ensemble with the Gibbs temperatures defined by
\begin{align}
\langle\beta^{\dag}_{s}\beta_s\rangle_G={\rm Tr} \left(\beta^{\dag}_{s}\beta_{s}{\hat \rho}_G\right).
\end{align}
In Ref.~\cite{HYL}, the quantum number $s$ corresponds to momentum $\cp k$. In the following, we will show that for a spin-orbit coupled BEC, the steady-state of the quench process can also be well described by a generalized Gibbs ensemble. Different from the previous cases, under SOC, due to the existence of two helicity branches in the excitation spectrum, we should now define two branches of momentum-dependent Gibbs temperatures.

In our case, the density operator can be written as
\begin{equation}
\hat{\rho}_{G}=Z_{G}^{-1}\exp(-\frac{1}{2}\sum_{\cp k}h_{\cp k}),\label{eqn:rg}
\end{equation}
where
\begin{align}
h_{\cp k}=\sum_{\lambda=\pm;\sigma=\uparrow,\downarrow}\frac{E_{\cp q+\lambda\cp k,\sigma}\beta^{\dag}_{\cp q+\lambda\cp k,\sigma}\beta_{\cp q+\lambda\cp k,\sigma}}{k_B T_{\lambda\cp k,\sigma}}.
\end{align}
Here, $T_{\cp k,\sigma}$ are the two branches of momentum dependent Gibbs temperature, and $E_{\cp q\pm \cp k,\sigma}$ is the quasi-particle spectrum in the steady state. With these, it is straightforward to derive the expression for the momentum distribution of quasi-particles in the generalized Gibbs ensemble
\begin{align}
\langle \beta^{\dag}_{\cp q+\cp k,\sigma}\beta_{\cp q+\cp k,\sigma}\rangle_G=\frac{A_{\cp k,\sigma}}{1-A_{\cp k, \sigma}},
\end{align}
where $A_{\cp k,\sigma}=\exp(-E_{\cp q+\cp k,\sigma}/k_B T_{\cp k,\sigma})$. And the Gibbs temperatures can be expressed as
\begin{align}
T_{\cp k,\sigma}=-\frac{E_{\cp q+\cp k,\sigma}}{k_B \ln A_{\cp k,\sigma}}.
\end{align}
From these, it is straightforward to derive the expression for the momentum distribution of atoms in the steady state (see Appendix for details).

\begin{figure}[tbp]
\centering
\includegraphics[width=8cm]{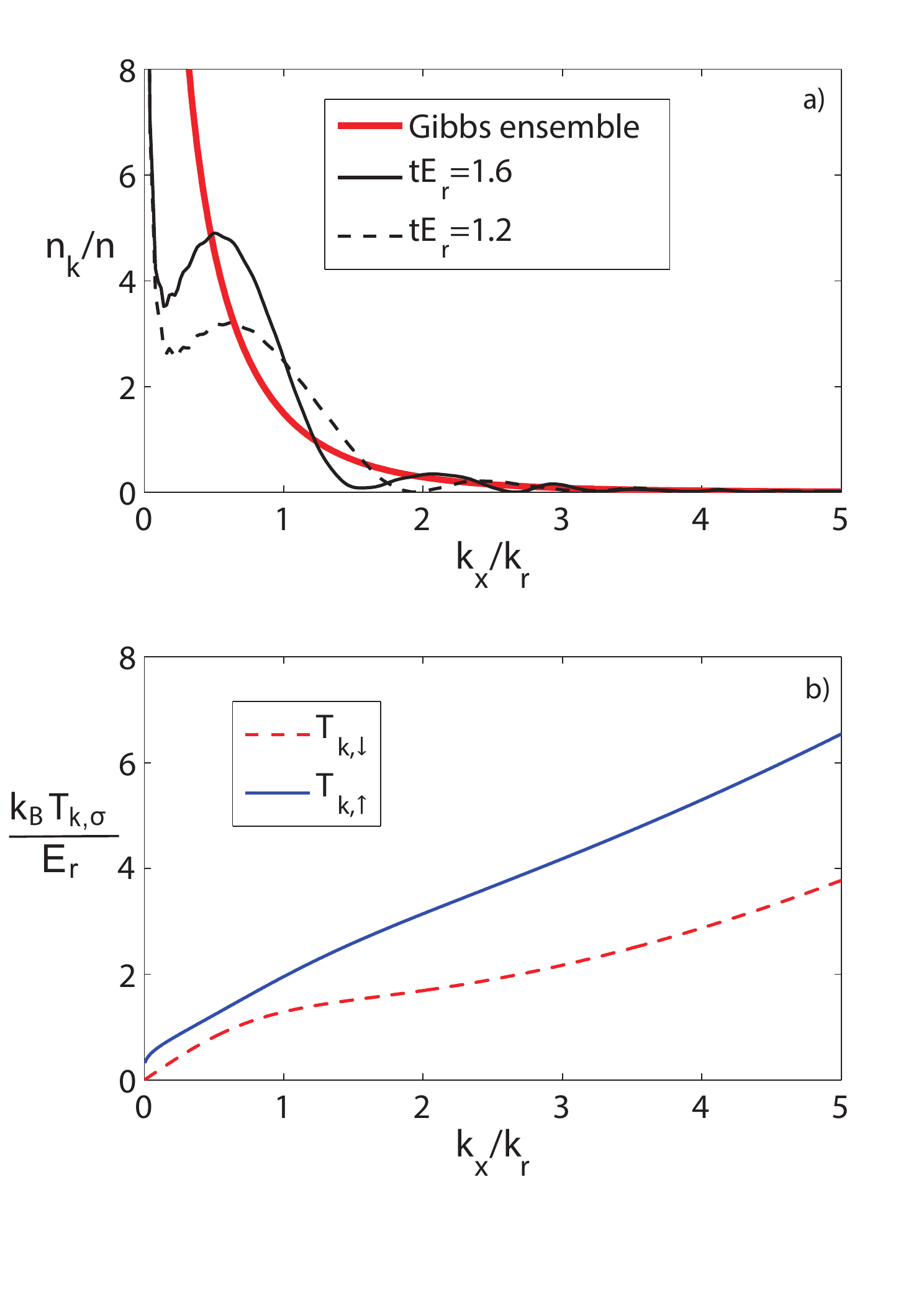}
\caption{(a) Momentum distribution of the Bose gas at different times after a change of interactions from an initial scattering length of $k_ra_i=0.01$ to $k_ra_f=0.2$. Thick red solid line is the momentum distribution given by the Gibbs ensemble. (b) Two branches of the momentum-dependent Gibbs temperature. Here, $\Omega/E_r=6$, $\delta=0$.}
\label{fig:fig4}
\end{figure}

\begin{figure}[tbp]
\centering
\includegraphics[width=8cm]{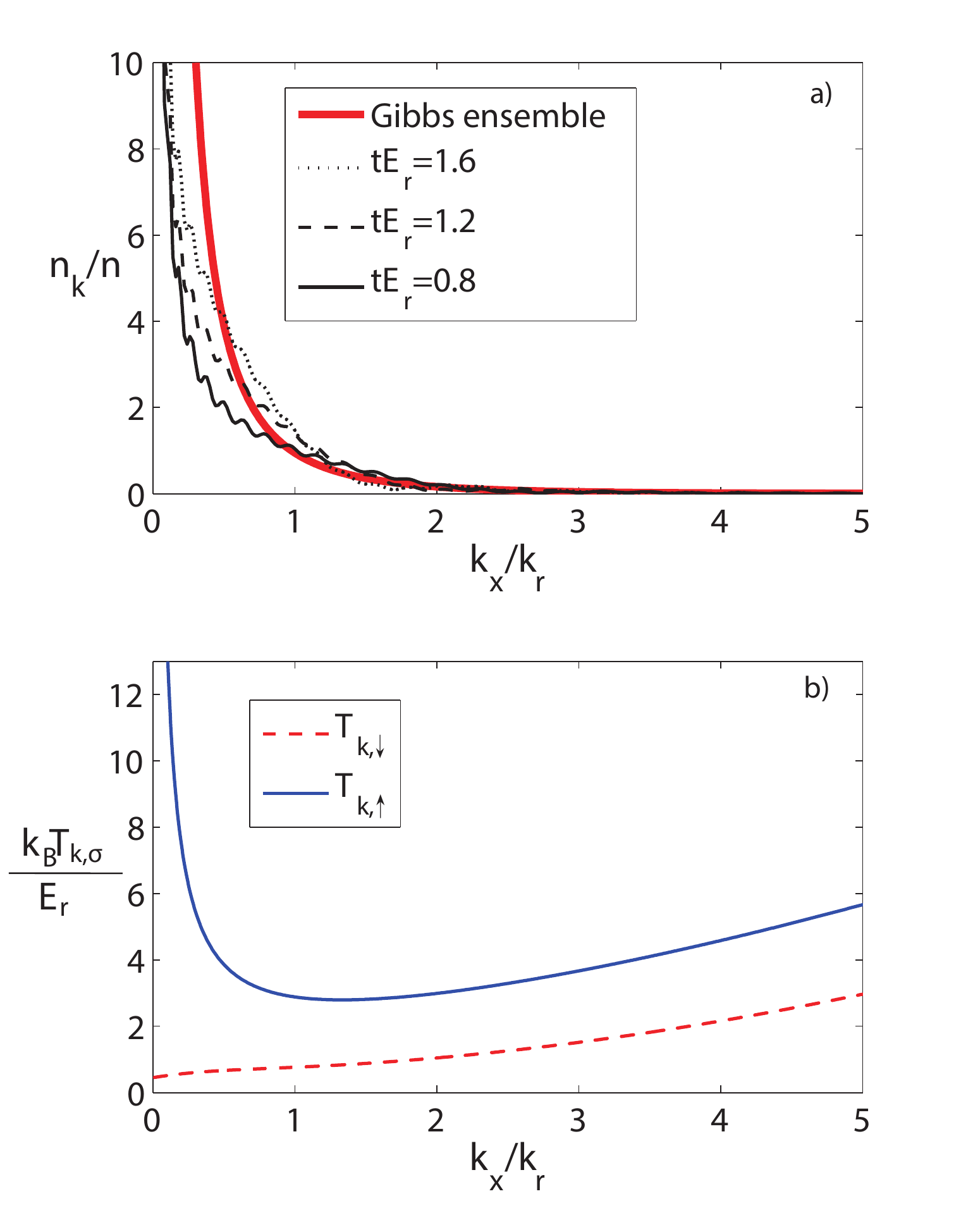}
\caption{(a) Momentum distribution of the Bose gas at different times after a change of Rabi frequency from $\Omega_i/E_r=2$ to $\Omega_f/E_r=6$. Thick red solid line is the momentum distribution given by the Gibbs ensemble. (b) Two branches of the momentum-dependent Gibbs temperature. Here, we have assumed SU(2)-invariant interactions with $k_r a_s=0.16$, and $\delta=0$.}
\label{fig:fig5}
\end{figure}

In Figs.~\ref{fig:fig4} and \ref{fig:fig5}, we compare the momentum distribution of the Bose gas at different times against that given by the generalized Gibbs ensemble. Although oscillations persist in the momentum distribution at long times, it is apparent that as the evolution time becomes longer, the center of the oscillations gradually approaches the distribution given by the generalized Gibbs ensemble. Therefore, in the long-time limit, the time average of the momentum distribution of atoms should be well described by the generalized Gibbs ensemble, just like the case in the absence of SOC. We also note that in Figs.~\ref{fig:fig4} and \ref{fig:fig5}, deviations from the distribution given by the generalized Gibbs ensemble exist mostly at small momenta. This can be attributed to the finite-time evolutions conducted in our numerical calculations. Indeed, as the evolution time increases, such deviations appear to become smaller. Importantly, as we have discussed previously, due to the existence of two helicity branches in the quasi-particle excitation spectrum, we now have to evoke two branches of Gibbs temperatures.

\section{Conclusion}

We study the quench dynamics of a spin-orbit coupled BEC using a self-consistent Bogoliubov theory. We consider cases where either the interaction strength or the SOC parameter undergoes sudden changes. In both cases, the quench dynamics typically leads to a steady state, in which the condensate fraction saturates while the momentum distribution of the atoms undergo fast oscillations. We demonstrate that while the momentum distribution in the steady state oscillates in the long-time limit, its average should converge to the distribution dictated by a generalized Gibbs ensemble, whose properties can be characterized by two branches of momentum-dependent Gibbs temperatures.

For a spin-orbit coupled BEC, the many-body ground state depends on the interaction strengths. As the relative magnitude of the intra- and inter-particle interaction changes, the ground state can either be in a plane-wave state or a spatially inhomogeneous stripe phase. For simplicity, we have chosen the plane-wave state as the initial state before the quench. It is straightforward, though numerically much more demanding to study the quench dynamics starting from a stripe phase. While we focus on sudden quenches in this work, a promising future direction is to study quench
processes of finite durations. For example, in these `slow' quenches, one may explore the Kibble-Zurek mechanism across second-order phase boundaries on the phase diagram of a spin-orbit coupled BEC~\cite{kb1,kb2}. It is also possible in principle to apply our approach to study the quench dynamics
of BEC under designed external potentials, where interesting dynamics have been reported very recently~\cite{sherman}.

\section*{Acknowledgements}
We thank H.-Y. Ling for helpful comments and discussions. This work is supported by NFRP (2011CB921200, 2011CBA00200), NKBRP (2013CB922000), NSFC (60921091, 11274009, 11374283, 11434011, 11522436, 11522545), and the Research Funds of Renmin University of China (10XNL016, 16XNLQ03). W. Y. acknowledges support from the ``Strategic Priority Research Program(B)'' of the Chinese Academy of Sciences, Grant No. XDB01030200.

\appendix
\section{Characterizing the steady state with the generalized Gibbs ensemble}

In this Appendix, we present in detail the derivation of the momentum distribution as well as the Gibbs temperatures for a spin-orbit coupled BEC in a general Gibbs ensemble. The temperature $T_{\cp k,\sigma}$ corresponding to each mode is determined by
\begin{equation}
\label{treq}
\langle\beta_{\cp q+\cp k,\sigma}^{\dagger}(t)\beta_{\cp q+\cp k,\sigma}(t)\rangle_G
=\langle\beta_{\cp q+\cp k,\sigma}^{\dagger}(t)\beta_{\cp q+\cp k,\sigma}(t)\rangle
,\end{equation}
where $\langle\beta_{\cp q+\cp k,\sigma}^{\dagger}(t)\beta_{\cp q+\cp k,\sigma}(t)\rangle_G$ has the form of $Tr(\beta_{\cp q+\cp k,\sigma}^{\dagger}\beta_{\cp q+\cp k,\sigma}\hat{\rho}_{G})$.
As what we have stated in Sec. II, the creation and annihilation operator of Bogoliubov quasi-particles should be written as
\begin{equation}
\left[\begin{array}{c}
\beta_{\cp q+\cp k,\uparrow}(t)\\
\beta_{\cp q+\cp k,\downarrow}(t)\\
\beta_{\cp q-\cp k,\uparrow}^{\dagger}(t)\\
\beta_{\cp q-\cp k,\downarrow}^{\dagger}(t)
\end{array}\right]=\tilde{E_u}U(0+)^{-1}U(0-)\left[\begin{array}{c}
\beta_{\cp q+\cp k,\uparrow}(0-)\\
\beta_{\cp q+\cp k,\downarrow}(0-)\\
\beta_{\cp q-\cp k,\uparrow}^{\dagger}(0-)\\
\beta_{\cp q-\cp k,\downarrow}^{\dagger}(0-)
\end{array}\right],
\end{equation}
in which
\begin{equation}
\tilde{E_u}=\left(\begin{array}{cccc}
e^{-i\int_{0}^{t}E_1dt} & 0 & 0 & 0\\
0 & e^{-i\int_{0}^{t}E_2dt} & 0 & 0\\
0 & 0 & e^{i\int_{0}^{t}E_3dt} & 0\\
0 & 0 & 0 & e^{i\int_{0}^{t}E_4dt}
\end{array}\right).
\end{equation}
The quasi-particle momentum distribution at any time can be calculated as
\begin{align}
\label{m1}
  \langle\beta_{\cp q+\cp k,\uparrow}^{\dagger}(t)\beta_{\cp q+\cp k,\uparrow}(t)\rangle & =|M_{13}|^{2}+|M_{14}|^{2}, \\
\label{m2}
  \langle\beta_{\cp q+\cp k,\downarrow}^{\dagger}(t)\beta_{\cp q+\cp k,\downarrow}(t)\rangle &  =|M_{23}|^{2}+|M_{24}|^{2},
\end{align}
where we define
\begin{equation}
  M=\left(\begin{array}{cccc}
M_{11} & M_{12} & M_{13} & M_{14}\\
M_{21} & M_{22} & M_{23} & M_{24}\\
M_{31} & M_{32} & M_{33} & M_{34}\\
M_{41} & M_{42} & M_{43} & M_{44}
\end{array}\right)=\tilde{E_u}U(0+)^{-1}U(0-).
\end{equation}
We then apply the definition in Eq.~\eqref{eqn:rg} and expand the right-hand side of Eq.~\eqref{treq} as
\begin{align}
\label{tr2}
&\langle\beta_{\cp q+\cp k,\sigma}^{\dagger}(t)\beta_{\cp q+\cp k,\sigma}(t)\rangle_G\nonumber\\
&=\frac{{\rm Tr}[\exp(-\frac{1}{2}\sum_{\cp k^{'}}h_{\cp k^{'}})\beta_{\cp q+\cp k,\sigma}^{\dagger}\beta_{\cp q+\cp k,\sigma}]}{{\rm Tr}[\exp(-\frac{1}{2}\sum_{\cp k^{'}}h_{\cp k^{'}})]},
\end{align}
Tracing out $\{\cp k,\sigma\}$ on all branches of the quasi-particle excitation spectrum, we have
\begin{align}
   & \langle\beta_{\cp q+\cp k,\sigma}^{\dagger}(t)\beta_{\cp q+\cp k,\sigma}(t)\rangle_G \notag\\
   &=\frac{\sum_{j=0}^{\infty}j A_j(\cp{k},\sigma)\prod_{(\cp{k}^{'},\sigma^{'})\neq(\cp{k},\sigma)} \sum_{j'=0}^{\infty}A_{j'}(\cp{k}^{'},\sigma^{'})}{\prod_{\cp{k}^{'},\sigma^{'}}\sum_{j=0}^{\infty}A_j(\cp{k}^{'},\sigma^{'})},
\end{align}
where $j=0,1,2,\cdots$, and
\begin{equation}
  A_j(\cp{k},\sigma)=\exp(-\frac{j E_{\cp{q}+\cp{k},\sigma}}{k_{B}T_{\cp{k},\sigma}}).
\end{equation}
After some simplification, we obtain
\begin{align}
\langle\beta_{\cp q+\cp k,\sigma}^{\dagger}(t)\beta_{\cp q+\cp k,\sigma}(t)\rangle_G=\frac{A(\cp k,\sigma)}{1-A(\cp k,\sigma)},
\end{align}
where
\begin{equation}
  A(\cp k,\sigma)=\exp(-\frac{E_{\cp q+\cp k,\sigma}}{k_{B}T_{\cp k,\sigma}}).
\end{equation}
Combining Eq.~\eqref{m1} ,Eq.~\eqref{m2} and Eq.~\eqref{treq}, we have
\begin{align}
\label{T1}
  T_{\cp k,\uparrow} & =\frac{E_{\cp q+\cp k,\uparrow}}{k_{B}\ln(\frac{1}{|M_{13}|^{2}+|M_{14}|^{2}}+1)}, \\
  \label{T2}
  T_{\cp k,\downarrow} & =\frac{E_{\cp q+\cp k,\downarrow}}{k_{B}\ln(\frac{1}{|M_{23}|^{2}+|M_{24}|^{2}}+1)}.
\end{align}\\
In the general Gibbs ensemble, the momentum distribution of atoms is
\begin{equation}
\label{ng}
n_{\cp{k}}(t)_{G}=\langle b_{\cp{q}+\cp{k},\uparrow}^{\dagger}(t)b_{\cp{q}+\cp{k},\uparrow}(t)+b_{\cp{q}+\cp{k},{\downarrow}}^{\dagger}(t)b_{\cp{q}+\cp{k},\downarrow}(t)\rangle_{G}.
\end{equation}
With Eq.~\eqref{b}, we rewrite Eq.~\eqref{ng} as
\begin{align}
\label{nG}
  n_{\cp k}(t)_{G}&=(|U_{11}|^{2}+|U_{21}|^{2})\langle\beta_{\cp q+\cp k,\uparrow}^{\dagger}\beta_{\cp q+\cp k,\uparrow}\rangle_{G}\notag\\
  &+(|U_{12}|^{2}+|U_{22}|^{2})\langle\beta_{\cp q+\cp k,\downarrow}^{\dagger}\beta_{\cp q+\cp k,\downarrow}\rangle_{G}\notag\\
  &+(|U_{13}|^{2}+|U_{23}|^{2})(\langle\beta_{\cp q-\cp k,\uparrow}^{\dagger}\beta_{\cp q-\cp k,\uparrow}\rangle_{G}+1)\notag\\
  &+(|U_{14}|^{2}+|U_{24}|^{2})(\langle\beta_{\cp q-\cp k,\downarrow}^{\dagger}\beta_{\cp q-\cp k,\downarrow}\rangle_{G}+1).
\end{align}
The results of the Gibbs temperatures Eq.~\eqref{T1}, Eq.~\eqref{T2} and the momentum distribution Eq.~\eqref{nG} are shown in Fig.~\ref{fig:fig4} and Fig.~\ref{fig:fig5}.

\end{document}